\documentclass[a4paper]{article}

\usepackage{INTERSPEECH2019}
\usepackage{multirow, siunitx}
\usepackage{diagbox}
\usepackage{makecell}
\usepackage{multirow}
\usepackage{arydshln}
\usepackage{rotating}
\usepackage{booktabs}
\usepackage{mathtools}
\usepackage[misc]{ifsym} 
\usepackage{bbding}

\usepackage[colorlinks,
            linkcolor=black,
            anchorcolor=black,
            citecolor=black,
            urlcolor=black]{hyperref}

\title{UNetGAN: A Robust Speech Enhancement Approach in Time Domain for Extremely Low Signal-to-noise Ratio Condition}
\name{Xiang Hao, Xiangdong Su*, Zhiyu Wang, Hui Zhang and Batushiren}
\address{
  Inner Mongolia Key Laboratory of Mongolian Information Processing Technology \\
  College of Computer Science, Inner Mongolia Univeristy, Hohhot, China}
\email{haoxiangsnr@gmail.com, cssxd@imu.edu.cn}

\begin{document}

\maketitle
\begin{abstract}
    Speech enhancement at extremely low signal-to-noise ratio (SNR) condition is a very challenging problem and rarely investigated in previous works. This paper proposes a robust speech enhancement approach (UNetGAN) based on U-Net and generative adversarial learning to deal with this problem. This approach consists of a generator network and a discriminator network, which operate directly in the time domain. The generator network adopts a U-Net like structure and employs dilated convolution in the bottleneck of it. We evaluate the performance of the UNetGAN at low SNR conditions (up to -20dB) on the public benchmark. The result demonstrates that it significantly improves the speech quality and substantially outperforms the representative deep learning models, including SEGAN, cGAN fo SE, Bidirectional LSTM using phase-sensitive spectrum approximation cost function (PSA-BLSTM) and Wave-U-Net regarding Short-Time Objective Intelligibility (STOI) and Perceptual evaluation of speech quality (PESQ).
\end{abstract}

\section{Introduction}

Speech enhancement is to separate the target speech from the background noise interference~\cite{se}.
It intends to improve the speech quality to optimize related signal processing systems, e.g., hearing prosthesis~\cite{se_for_hearing_aid}, mobile telecommunication~\cite{se_for_mobile}, and automatic speech recognition~\cite{se_for_asr}, and so on.
This problem has been widely studied for a long time, and a large number of methods were proposed~\cite{wang_2018_overview}~\cite{se_book_2018}.
However, few of these methods pay attention to speech enhancement at low signal-to-noise ratio (SNR) condition which is more important and challenging than that at high SNR condition.
There are many communication scenarios at extremely low SNR condition. For examples, workers communicating with the walkie-talkie in a metal cutting factory, mechanics communication with wireless headset when test a helicopter, and so on. 
Even in some noisy environments, people can only use gestures to communicate, because the sound collected by the microphone has serious noise, which makes the counterpart unable to hear clearly.
Speech enhancement for low SNR determines whether the speeches can be heard clearly and understood accurately while that at high SNR only makes the speech more comfortable for listeners.
From this point of view, speech enhancement at low SNR is more important than at high SNR. 

This paper integrates U-Net and dilated convolution operation~\cite{dialted_conv_first_used} into generative adversarial network (GAN)~\cite{goodfellow_generative_2014} based framework and proposes a speech enhancement approach (named UNetGAN) for extremely low SNR condition.
Our motivation is as follows. At first, speech enhancement can be seen as a special type of speech separation, which separates the clean speech from the mixture.
Among the deep learning based models of speech separation, Wave U-Net~\cite{wave_u_net}, a U-Net operates in the time domain, achieves state-of-the-art performance while requires no preprocessing.
Second, GAN can improve the performance of the generator network through play a min-max game between the generator network and discriminator network. 
Its effectiveness in speech enhancement is also proved in the work~\cite{wang_gan}.
Third, dilated convolution can expand the receptive field size and take large temporal contexts into account~\cite{deeplab}.

The proposed UNetGAN consists of a generator network and a discriminator network, which operates in the time domain and is trained in an adversarial way.
The generator network adopts a U-Net like structure, in which dilated convolution is used between the downsampling and upsampling blocks.
The discriminator network is a conventional convolution neural network, involving batch normalization~\cite{batch_norm} and leaky rectified linear unit (Leaky ReLU)~\cite{leaky_relu}.
Once trained, the generator can be used for speech enhancement.

Our approach is evaluated at extremely low SNR conditions (up to -20dB) on the public dataset.
Comparison is also made between the proposed approach and other representative speech enhancement approaches based on deep learning, including SEGAN~\cite{segan}, cGAN for SE~\cite{cgan_se}, bidirectional LSTM using phase-sensitive spectrum approximation cost function (PSA-BLSTM)~\cite{phase_sensitive_bilstm}, and Wave-U-Net~\cite{wave_u_net}.
Experiment results demonstrate that our approach significantly improves the speech quality and substantially outperforms other approaches in terms of Short-Time Objective Intelligibility (STOI)~\cite{STOI} and Perceptual evaluation of speech quality (PESQ)~\cite{PESQ}.

Although Pascual et al.~\cite{segan} first used GAN in the time domain (named SEGAN) for speech enhancement, there are two critical differences between SEGAN and our approach.
First, the network structure in our approach is different from that in SEGAN.
Second, our approach operates directly in the time domain while SEGAN applies a high-frequency preemphasis filter to the input data.
There is another GAN-based speech enhancement approach is proposed by Michelsanti et al.~\cite{cgan_se}. Except the difference in network structure, it operates in the frequency domain while our approach operates in the time domain.
Our approach also differ from the following speech enhancement approaches, which are based on fully convolutional neural network~\cite{full_conv_e2e_2018}~\cite{waveform_baesed_se_by_full_conv}~\cite{e2e_full_conv_time} and U-Net~\cite{song_seperation_unet}~\cite{rnn_unet_in_tf_domain}~\cite{a_new_framework_wang}.
We introduce U-Net into GAN-based architecture and takes adversarial learning to train it.
It also means the loss function of the U-Net in our approach are different from that in the above approaches.
\begin{figure*}[!ht]
    \centering
    \centerline{\includegraphics[width=160mm]{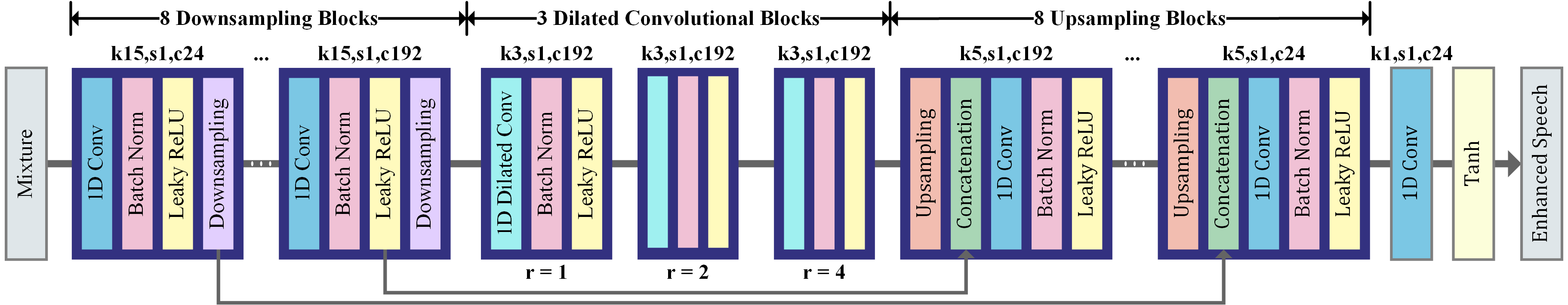}}
    \caption{
        The generator of UNetGAN.
        $k$, $s$ and $c$ represent kernel size, stride and channel number of 1D convolution, respectively. $r$ represents dilated rate of 1D dilated convolution.
    }
    \label{fig:G_Arch}
\end{figure*}

There are two main advantages of the proposed approach.
First and foremost, our approach significantly improves the speech quality and achieves state-of-the-art performance at extremely low SNR conditions.
Second, our approach employs adversarial learning to improve the U-Net in the time domain and applies dilated convolution to enlarge the receptive field size.
Third, our approach works end-to-end. %

\begin{figure}
    \centering
    \centerline{\includegraphics[width=7.2cm]{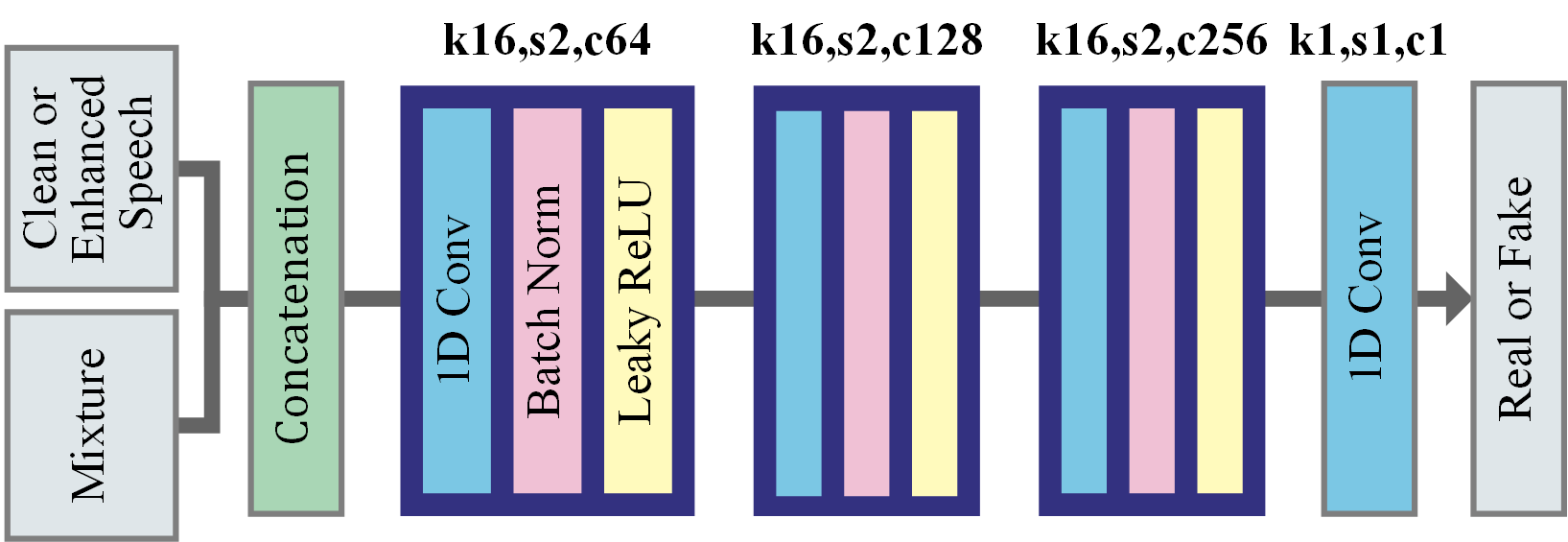}}
    \caption{
        The discriminator of UNetGAN.
        $k$, $s$ and $c$ represent kernel size, stride and channel number of 1D convolution, respectively.
    }
    \label{fig:D_Arch}
\end{figure}
\section{Approach}
\subsection{Architecture}
This paper proposes a conditional GAN (cGAN)~\cite{cgan} based approach to perform speech enhancement, which consists of two components: a generator network and a discriminator network. The generator adopts U-Net like structure and dilated convolution operation. The discriminator is a conventional CNN, involving batch normalization and Leaky ReLU activation.
Both of these two networks operate directly in the time domain. %
The discriminator $D$ is trained to classify the speeches as either being from the training data (real, close to 1) or the generator $G$ (fake, close to 0).
It is conditioned on the mixture $x$ and maps the clean speech $y$ or the enhanced speech $\hat{y}$ to the real data distribution: $D(x, y)$ or $D(x, \hat{y}) \rightarrow (0, 1)$.
The generator $G$ maps from the mixture $x$ to the enhanced speech $\hat{y}$: $G(x) \rightarrow \hat{y}$, trying to confuse the discriminator $D$.
They play a min-max game.
The objective can be expressed as:
\begin{equation}
    \text{arg} \,\underset{G}{\text{min}}\, \underset{D}{\text{max}}( \mathbb{E}_{x, y}[logD(x, y)] \nonumber  +  \mathbb{E}_{x, \hat{y}}[log(1 - D(x, \hat{y})])
\end{equation}
where G tries to minimize this objective against the adversarial D tries to maximize it.

The detailed structure of the generator is shown in Fig.~\ref{fig:G_Arch}.
The downsampling (DS) and the upsampling (US) parts are similar to that described in~\cite{wave_u_net} except several dilated convolution blocks are added between them.
At first, the mixture $x$ is converted into an increasing number of higher-level features on coarser time scales using a series of DS blocks.
Next, these features are processed by three successive dilated convolution blocks to incorporate larger context.
Subsequently, the features are combined with the earlier local, high-resolution features using US blocks through skip connection, yielding multi-scale features which are used for making predictions.

The DS blocks in the generator have eight levels in total. Each successive level operates at half the time resolution as the previous one while the number of channels increased gradually at an interval of twenty-four.
Each DS block here takes 1D convolution, followed by batch normalization, Leaky ReLU and downsampling.
The parameters of 1D convolution can be found above each DS block in Fig.~\ref{fig:G_Arch}, in which $k$, $s$ and $c$ represent kernel size, stride and channel number of 1D convolution, respectively.
It performs same padding to produce the output of the same time resolution as the input.
Batch normalization is used to ensure network performance and stability.
We use Leaky ReLU as activation function except for the final layer, which uses Tanh.
Decimate discards features for every other time step to halve the time resolution.
In the dilated convolutional blocks, we use three successive dilated convolution operations with different dilated rate ($r = 1, 2, 4$) to gradually extract the features at the desirable resolutions.
The resulting feature maps are used as the input of the US blocks.
The details of dilated convolution are described in Section~\ref{dialted_conv}.
In the US blocks, upsampling performs linear interpolation in the time direction by a factor of two.
The channel number at each level decrease at an interval of twenty-four.

The discriminator is illustrated in Fig.~\ref{fig:D_Arch}.
The clean speech or the enhanced speech is concatenated with the mixture, and transformed to an increasing number of feature maps using 1D convolution, batch normalization and Leaky ReLU.
After the three convolution blocks, the feature maps eventually are compressed into a high-level representation.

These two networks are trained alternately. For the fixed generator $G$, the discriminator $D$ is trained to distinguish between the clean speeches and the enhanced speeches.
When the discriminator is optimal, it can be frozen and the generator $G$ can continue to be trained to reduce the accuracy of the discriminator.

\subsection{Dilated Convolution}
\label{dialted_conv}
As mentioned above, the dilated convolution is used in the generator network.
This operation was originally developed for wavelet transform~\cite{atrous_algorithm} and later called dilated convolution in~\cite{dialted_conv_first_used}.
It inflates the kernel by inserting spaces between the kernel elements.
It allows us to enlarge the receptive field size to incorporate larger context~\cite{understanding_dilated_conv}.
For a 1-D input signal $x[i]$, the output $y[i]$ of dilated convolution on it with a filter $w[k]$ of length $K$ is defined as:
\begin{equation}
    y[i] = \sum_{k=1}^{K} x[i + r \cdot k]w[k]
\end{equation}
where $r$ is the dilated rate. When $r = 1$, the dilated convolution is the same as the conventional convolution.

\begin{figure}
    \centering
    \centerline{\includegraphics[width=7.5cm]{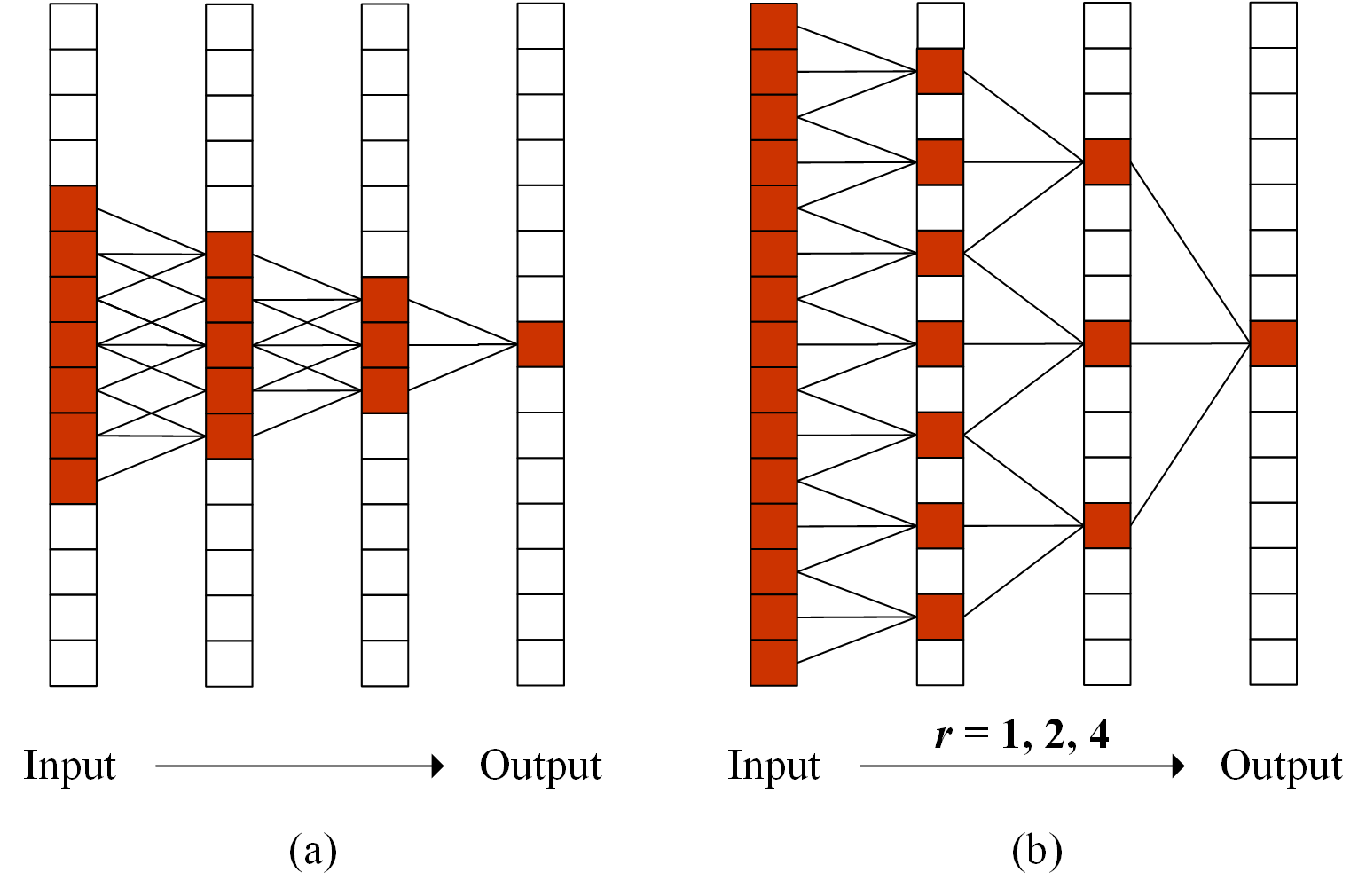}}
    \caption{
        (a) A three layers CNN using 1D conventional convolution operation.
        (b) A three layers CNN using 1D dilated convolution operation with exponentially increasing dilated rate ($r = 1, 2, 4$). %
    }
    \label{fig:dilation}
\end{figure}

Figure~\ref{fig:dilation} illustrates the conventional convolution and the dilated convolution ($r= 1, 2, 4$) on 1-D signals, in which $stride=1$, $kernel~size=3$. 
Fig.~\ref{fig:dilation}(a) shows that the receptive field size is seven after the three sequential convolution operations, which is linear with the number of layers. 
When using exponentially increasing dilated rates ($r = 1, 2, 4$), as in Fig.~\ref{fig:dilation}(b), the receptive field size exponentially increases to 15.
The dilated convolution blocks in our model use the same exponentially increasing dilated rate ($r = 1, 2, 4$) as Fig.~\ref{fig:dilation}(b), which ensures that the size of receptive field increases exponentially.

\subsection{Loss Function}
The task of training a neural network is one of finding good parameters in an iterative process, in which the loss function plays a central role through gradient backpropagation for parameter adjustments.
Previous works about U-Net~\cite{wave_u_net}~\cite{biomedical_image_segmentation_UNet} use mean-square error (MSE) as the loss function and obtain good performances, proving the effectiveness of MSE.
GAN-based networks usually employ adversarial loss as the loss function to ensure the generated results in line with expectations.
The generator gradually improves with the decrease of the adversarial loss.
Our approach integrates MSE and adversarial loss as the loss function of the generator network, as Eq.~\ref{eq:G_loss}.
\begin{equation}
    \mathcal{L}_{G} = \underbrace{\mathbb{E}_{x}[log (1 - D(x, G(x)))]}_{Adversarial~Loss} + \lambda \underbrace{\mathbb{E}_{x, y} [y - G(x)]^2}_{MSE}
    \label{eq:G_loss}
\end{equation}
where $x$ is the mixture, $y$ is the clean speech in the training dataset, and $\lambda$ is the weight of MSE which is determined through the validation experiment.

The discriminator is still unchanged.
That is, it is responsible for mapping the clean speech $y$ to true and the enhanced speech $G(x)$ to false conditioned on the mixture $x$.
Thus, its loss function can be written as Eq.~\ref{eq:D_loss}.
\begin{equation}
    \mathcal{L}_{D} = - \mathbb{E}_{x}[log(1 - D(x, G(x)))] - \mathbb{E}_{x, y}[log D(x, y)]
    \label{eq:D_loss}
\end{equation}

\section{Experiment}
\subsection{Dataset and metrics}
To evaluate the proposed approach, the TIMIT corpus~\cite{TIMIT} and NOISEX-92 corpus~\cite{NoiseX92} are used in the experiment.
The TIMIT corpus is used as the clean database and the NOISEX-92 corpus is used as interference.
We randomly selected \num{750} utterances from the TIMIT and divided them into three parts: the training part (\num{600} utterances), the validation part (\num{50} utterances) and the test part (\num{100} utterances).

With respect to the training set, we selected babble, factoryfloor1, destroyerengine and destroyerops from NOISEX-92 corpus.
The first 2 minutes of each noise are mixed with the clean speech in the training part at one of 4 SNRs (0dB, -5dB, -10dB, -15dB).
In total, this yields \num{9600} training samples, each of which consists of a mixture and its corresponding clean speech.
Beside the noises in the training set, we selected factoryfloor2 (from NOISEX-92 corpus) to evaluate generalization performance.
The last 2 minutes of each noise are mixed with the test utterances at one of 9 SNRs (0dB, -3dB, -5dB, -7dB, -10dB, -12dB, -15dB, -17dB, -20dB), resulting in \num{4500} test samples.
The validation set is built in the same way as the test set, which includes \num{2250} samples.
The sampling rate of all samples is \num{16000} Hz. 
The noise is divided into two sections to ensure that the test noise is not repeated in the training set.

We use STOI and PESQ score to measure speech intelligibility and quality, respectively.

\subsection{Training}

As mentioned, each sample consists of a mixture and its corresponding clean speech. 
All samples vary in length.
In each training epoch, we randomly sample continuous 16384 time frames from the mixture and the clean speech respectively, and input them into the networks of UNetGAN.
We use the Adam optimizer~\cite{adam} with $\text{leaning rate}=0.0002$, decay rates $\beta_{1} = 0.9$ and $\beta_{2} = 0.999$.
We set batch size to \num{150} and negative slope of Leaky ReLU to \num{0.1}.
The $\lambda$ in Eq.~\ref{eq:G_loss} is set according to the validation experiment, which equals \num{20} in the best model.

We perform gradient descent on the generator $G$ and the discriminator $D$ alternatively to optimize both of them.
Generator loss shows some oscillatory behavior initially and gradually converges after \num{900} epochs.
It approximates to $0.771$ eventually.

\begin{table*}[!htb]
    \caption{
        STOI and PESQ of UNetGAN at different SNRs and different noises.
        N1, N2, N3, N4, N5 are babble, factoryfloor1, destroyerengine, destroyerops and factoryfloor2, respectively.
    }
    \label{diff_noise}
    \centering
    \scriptsize
    \renewcommand\arraystretch{1}
    \setlength{\tabcolsep}{1.2mm}{
        \begin{tabular}{cccccccccccccccccccc}
            \toprule
            \multirow{3}{*}{Noise} & \multirow{3}{*}{Target} & \multicolumn{9}{c}{STOI}  & \multicolumn{9}{c}{PESQ}                                                                                                                                                                              \\
            \cmidrule(r){3-11}\cmidrule(lr){12-20}
                                   &                         & \multicolumn{4}{c}{Seen} & \multicolumn{5}{c}{Unseen} & \multicolumn{4}{c}{Seen} & \multicolumn{5}{c}{Unseen}                                                                                                                 \\
            \cmidrule(r){3-6}\cmidrule(r){7-11}\cmidrule(lr){12-15}\cmidrule(r){16-20}
                                   &                         & 0dB                       & -5dB                        & -10dB                     & -15dB                       & -3dB    & -7dB    & -12dB   & -17dB   & -20dB   & 0dB     & -5dB    & -10dB   & -15dB   & -3dB    & -7dB    & -12dB   & -17dB   & -20dB   \\
            \midrule
            \multirow{2}{*}{N1}    & Mixture                 & 0.735                     & 0.635                       & 0.531                     & 0.440                       & 0.676 & 0.593 & 0.492 & 0.412 & 0.378 & 1.863 & 1.483 & 1.158 & 0.915 & 1.629 & 1.343 & 1.049 & 0.874 & 0.833 \\
                                   & Enhanced                & 0.903                     & 0.879                       & 0.840                     & 0.780                       & 0.890 & 0.866 & 0.818 & 0.749 & 0.695 & 2.659 & 2.467 & 2.226 & 1.975 & 2.550 & 2.380 & 2.129 & 1.870 & 1.722 \\
            \midrule
            \multirow{2}{*}{N2}    & Mixture                 & 0.755                     & 0.650                       & 0.551                     & 0.475                       & 0.692 & 0.609 & 0.517 & 0.452 & 0.426 & 1.783 & 1.507 & 1.309 & 1.188 & 1.609 & 1.405 & 1.249 & 1.163 & 1.132 \\
                                   & Enhanced                & 0.897                     & 0.874                       & 0.841                     & 0.793                       & 0.884 & 0.862 & 0.824 & 0.766 & 0.716 & 2.517 & 2.339 & 2.146 & 1.918 & 2.413 & 2.266 & 2.063 & 1.811 & 1.651 \\
            \midrule
            \multirow{2}{*}{N3}    & Mixture                 & 0.755                     & 0.665                       & 0.566                     & 0.477                       & 0.703 & 0.626 & 0.528 & 0.449 & 0.415 & 1.935 & 1.540 & 1.190 & 0.928 & 1.696 & 1.390 & 1.078 & 0.867 & 0.762 \\
                                   & Enhanced                & 0.910                     & 0.888                       & 0.853                     & 0.802                       & 0.898 & 0.876 & 0.835 & 0.775 & 0.723 & 2.751 & 2.572 & 2.354 & 2.091 & 2.651 & 2.485 & 2.252 & 1.983 & 1.810 \\
            \midrule
            \multirow{2}{*}{N4}    & Mixture                 & 0.721                     & 0.610                       & 0.507                     & 0.432                       & 0.654 & 0.566 & 0.473 & 0.411 & 0.390 & 1.775 & 1.392 & 1.100 & 0.859 & 1.547 & 1.262 & 0.963 & 0.820 & 0.792 \\
                                   & Enhanced                & 0.897                     & 0.869                       & 0.830                     & 0.774                       & 0.882 & 0.855 & 0.810 & 0.745 & 0.691 & 2.656 & 2.458 & 2.230 & 1.980 & 2.541 & 2.368 & 2.131 & 1.869 & 1.701 \\
            \midrule
            \multirow{2}{*}{N5}    & Mixture                 & 0.830                     & 0.755                       & 0.664                     & 0.567                       & 0.787 & 0.720 & 0.625 & 0.530 & 0.481 & 2.231 & 1.851 & 1.487 & 1.183 & 2.004 & 1.701 & 1.354 & 1.083 & 0.972 \\
                                   & Enhanced                & 0.876                     & 0.813                       & 0.708                     & 0.575                       & 0.843 & 0.775 & 0.657 & 0.540 & 0.500 & 2.502 & 2.179 & 1.806 & 1.425 & 2.313 & 2.030 & 1.653 & 1.285 & 1.137 \\
            \bottomrule
        \end{tabular}
    }
\end{table*}

\subsection{Baseline Approaches}
This paper compares UNetGAN with following approaches, including GAN-based approach in the time domain (SEGAN~\cite{segan}), GAN-based approach in the frequency domain (cGAN for SE~\cite{cgan_se}), BiLSTM-based approach using time-frequency mask (PSA-BLSTM~\cite{phase_sensitive_bilstm}) and U-Net based approach in the time domain (Wave-U-Net~\cite{wave_u_net}).
\begin{itemize}
    \item SEGAN: SEGAN is a GAN-based speech enhancement approach, which operates at the waveform level. It applies a high-frequency preemphasis filter to all input speeches. We use the same implementation~\cite{segan_implement_github} as SEGAN and keep its default parameters unchanged.

    \item cGAN for SE: cGAN for SE make use of the pix2pix framework proposed by Isola et al.~\cite{pix2pix} to learn a mapping from the spectrogram of noisy speech to the enhanced counterpart. We adopt the same preprocessing method and implementation as mentioned in~\cite{cgan_se}.

    \item PSA-BLSTM: PSA-BLSTM is a bidirectional LSTM network for speech enhancement, using a phase-sensitive spectrum approximation (PSA) cost function. We reimplement the model and use the same hyperparameter as that in~\cite{phase_sensitive_bilstm}. %

    \item Wave-U-Net: Wave-U-Net is an adaptation of the U-Net architecture to the one-dimensional time domain to perform end-to-end audio source separation. We use the implementation on~\cite{wave_u_net_implement_github} and its default parameters.
\end{itemize}

\subsection{Result and Discussion}
\begin{table}
    \centering
    \caption{The average performances of UNetGAN and the baseline approaches on the test set.} %
    \label{compare_with}
    \scriptsize
    \renewcommand\arraystretch{1.2}
    \setlength{\tabcolsep}{6mm}{
        \begin{tabular}{lrr}
            \toprule
            Method      & STOI           & PESQ           \\
            \midrule
            Mixture     & 0.576          & 1.317          \\
            SEGAN       & 0.586          & 1.303          \\
            cGAN for SE & 0.590          & 1.220          \\
            PSA-BLSTM   & 0.646          & 1.820          \\
            Wave-U-Net  & 0.654          & 1.584           \\
            UNetGAN     & \textbf{0.802} & \textbf{2.140} \\
            \bottomrule
        \end{tabular}
    }
\end{table}
\begin{figure}
    \centering
    \centerline{\includegraphics[width=8.2cm]{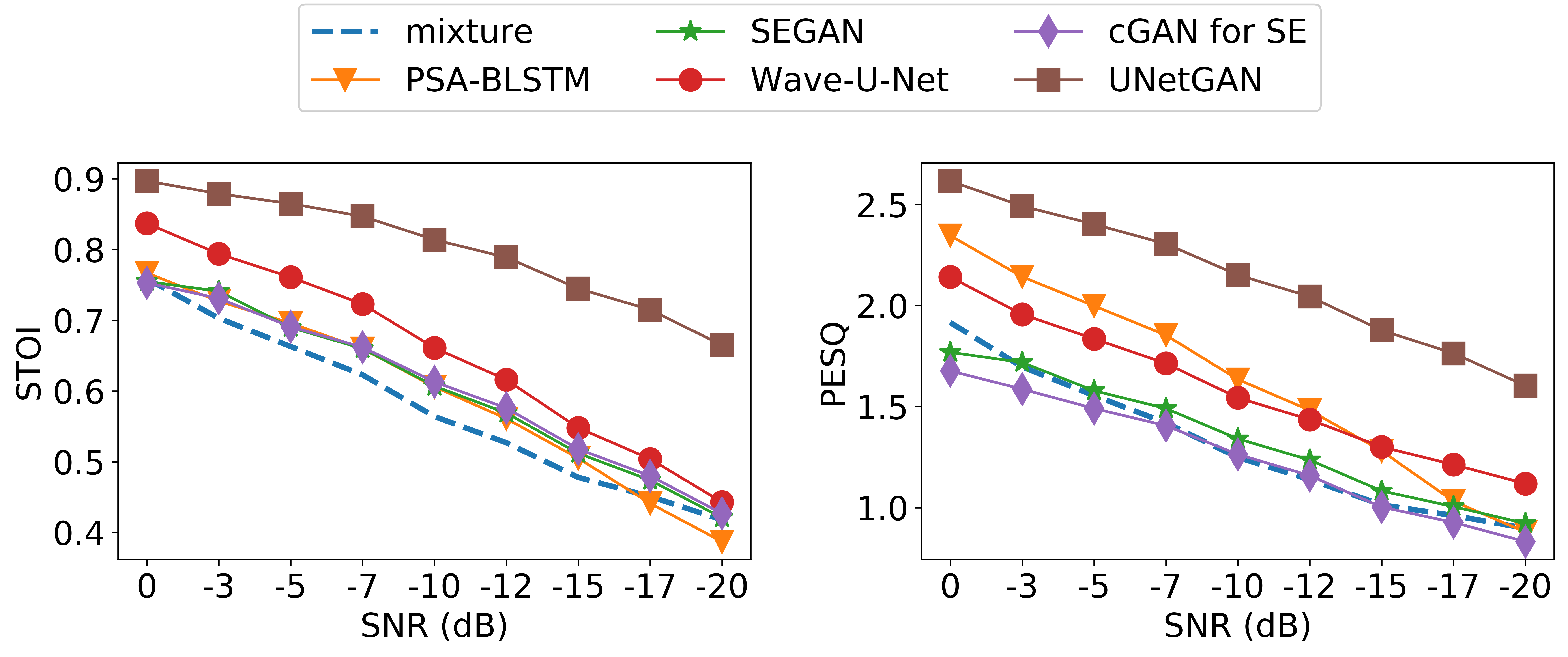}}
    \caption{
        Illustration of the average performances of UNetGAN and the baseline approaches at different SNRs.}
    \label{fig:compare}
\end{figure}
Table~\ref{diff_noise} presents the enhanced speech and the mixture from 0dB to -20dB in terms of STOI and PESQ.
The ``Mixture" lines in the table represent the mixture, and the ``Enhanced" lines represent the enhanced speeches using UNetGAN.
The ``Seen" columns mean the SNR conditions exist in the training set while the ``Unseen" columns indicate the SNR conditions that do not exist in the training set. 
The noise N1, N2, N3, N4, N5 are babble, factoryfloor1, destroyerengine, destroyerops and factoryfloor2, respectively.
Among them, factoryfloor2 is not used in the model training section.

In Table~\ref{diff_noise}, there are significant improvements of STOI and PESQ after speech enhancement at all conditions. 
The SNRs lower than -5dB are generally considered as extremely low SNR conditions.%
The average improvement on STOI and PESQ are \SI{39.16}{\percent} and \SI{62.55}{\percent}, respectively. 
This indicates that our approach is very effective.
At ``Seen" SNR conditions, the average improvement on STOI and PESQ are \SI{34.74}{\percent} and \SI{57.80}{\percent}.
The corresponding improvments at ``Unseen" conditions are \SI{43.16}{\percent} and \SI{67.01}{\percent}, reflecting the stability of UNetGAN. 
The average improvement on unseen noise ``factoryfloor2" achieves a \SI{5.55}{\percent} STOI and a \SI{18.33}{\percent} PESQ, proving that UNetGAN possess a good generalization ability. 

Table~\ref{compare_with} lists the performances of UNetGAN and the baseline approaches on the test set. 
Among the baseline approaches, Wave-U-Net gets the highest STOI (\SI{12.54}{\percent} higher than the mixture) and PSA-BLSTM obtains the highest PESQ (\SI{31.13}{\percent} higher than the mixture). 
UNetGAN gains a \SI{39.16}{\percent} improvement on STOI and a \SI{62.55}{\percent} improvement on PESQ comparing with the mixture. 
It is obviously that UNetGAN performs far better than any other baseline approach.
Although both Wave-U-Net and UNetGAN use U-Net like structures, the dilated convolution operation, adversarial learning and well-defined structure contribute to the excellent performance of UNetGAN.

Figure~\ref{fig:compare} illustrates STOI and PESQ of UNetGAN and the baseline approaches at different SNRs.
It suggests that UNetGAN performs much better than others at above all SNRs.
In Fig.~\ref{fig:compare}, as the SNR decreases, all the baselines gradually approach the mixture on STOI and PESQ.
It means that the effect of these approaches are very limited when SNR is very low. 
On the contrary, our method exhibits strong robustness at extremely low SNRs.

\section{Conclusion}
Speech enhancement at low SNR condition is a challenging task.
Concerning the performance of U-Net in voice separation and the effect of GAN in network training, this paper integrated U-Net into GAN-based framework and proposed a end-to-end speech enhancement approach for extremely low SNR condition. We also used dilated convolution operation to enlarge the receptive field size in feature extraction. 
Our approach achieves state-of-the-art performance and significantly outperforms SEGAN, cGAN for SE, PSA-BLSTM and Wave-U-Net. The average improvement on STOI and PESQ achieves \SI{39.16}{\percent} and \SI{62.55}{\percent} respectively. %
Experiment also proves that our models are robust to unseen low SNR conditions and noises.
To our best knowledge, this papar is the first time that explores speech enhancement at extremly low SNR conditions (up to $-20$dB).

\bibliographystyle{IEEEtran}
\bibliography{UNetGAN}

\end{document}